\documentclass[%
 reprint,
superscriptaddress,
showpacs,preprintnumbers,
 amsmath,amssymb,
 aps,
prb,
]{revtex4-1}
\usepackage{slashed}
\usepackage{graphicx}
\usepackage{dcolumn}
\usepackage{bm}

\begin{document}
\title{Quantum spin Hall effect in two-dimensional metals without spin-orbit coupling}

\author{Aiying Zhao}
\affiliation{Institute of Theoretical Physics, University of Science and Technology Beijing,  Beijing 100083, China}
\affiliation{Department of Physics, University of Central Florida, Orlando, FL 32816-2385, USA}
\author{Qiang Gu}
\email{qgu@ustb.edu.cn, corresponding author}
\affiliation{Institute of Theoretical Physics, University of Science and Technology Beijing,  Beijing 100083, China}
\author{Timothy J. Haugan}
\affiliation{U. S. Air Force Research Laboratory, Wright-Patterson Air Force Base, Ohio 45433-7251, USA}
\author{Thomas J. Bullard}
\affiliation{U. S. Air Force Research Laboratory, Wright-Patterson Air Force Base, Ohio 45433-7251, USA}
\author{Richard A. Klemm}
\email{richard.klemm@ucf.edu, corresponding author}
\affiliation{Department of Physics, University of Central Florida, Orlando, FL 32816-2385, USA}
\affiliation{U. S. Air Force Research Laboratory, Wright-Patterson Air Force Base, Ohio 45433-7251, USA}
\date{\today}

\baselineskip12pt


\begin{abstract}
The quantum spin Hall effect  has been observed in topological insulators using spin-orbit coupling  as the probe, but it has not yet been observed in a metal.
An experiment is proposed to measure the  quantum spin Hall effect of an electron or hole in a two-dimensional  (2D) metal by using  the previously unexplored but relativistically generated 2D quantum spin Hall Hamiltonian, but  without using spin-orbit coupling.  A long cylindrical solenoid lies normally through the inner radius of a 2D metallic Corbino disk. The current $I_S$ surrounding the solenoid produces an azimuthal magnetic vector potential but no magnetic field in the disk.  In addition, a radial electric field is generated across the disk by  imposing either (a) a potential difference $\Delta v$ or (b) a radial  charge current ${\bm I}$ across its inner and outer radii. Combined changes in $I_S$  and  in either $\Delta v$ or ${\bm I}$ generate spontaneously quantized azimuthal charge and spin currents.  The experiment is designed to measure these quantized  azimuthal charge and spin currents in the disk consistently. The quantum Hamiltonians for experiments (a) and (b)  are both solved exactly.    A  method to control the Joule heating is presented, which could potentially allow the quantum spin Hall measurements to be made at room temperature. Extensions of this design to an array of thermally-managed solenoids, each surrounded by thermally-managed stacks of 2D metallic Corbino disks, could function as a quantum computer that could potentially operate at room temperature.
\end{abstract}

 \vskip0pt
\maketitle

\section{Introduction}
In the hydrogen atom, one of the leading relativistic corrections to the non-relativistic limit is proportional to $({\bm p}\times{\bm E})\cdot{\bm\sigma}$ \cite{Dirac,FW,BjorkenDrell}, where ${\bm p}$ is the quantized momentum of the electron, the electric field ${\bm E}=-{\bm\nabla}\Phi$, where $\Phi$ is the radial electrostatic potential in spherical coordinates, and the components of ${\bm \sigma}$ are the Pauli matrices representing the electron spin. Since ${\bm E}||{\bm r}$ and  the angular momentum ${\bm L}={\bm r}\times{\bm p}$, such a term in the Hamiltonian  represents spin-orbit coupling, but  differs from the  classical  Hall effect in three-dimensional metals that results from  both an applied ${\bm E}$ and an applied magnetic induction ${\bm B}$, but does not include the particle's spin.

More recently, there has been a very large interest in the quantum spin Hall effect in thin topological insulators and insulating quantum wells \cite{Kane,Bernevig,Hasan,Qi,Wu1}.
In those studies, the model Hamiltonian  was also proportional to $({\bm p}\times{\bm E})\cdot{\bm\sigma}$, and since the electrons in such insulators only travel on or near to the sample surface, to first approximation, ${\bm E}||{\bm r}$, where the position ${\bm r}$ of the electron is measured from the center of the top surface under study. Moreover in  topological insulators, including that 1$T'$ form of monolayer WTe$_2$ studied with regard to the quantum spin Hall effect \cite{Wu1}, the Dirac-cone electronic dispersion locks the electron spins onto their momenta, and the protected edge currents are insensitive to backscattering from defects or from travelling around the corners of the top surface. In those experiments, the conductance was observed to be quantized in integral units of $e^2/h$, as for the ordinary quantum Hall effect \cite{Mahan}, where $e$ and $h$ are the electric charge and Planck's constant, respectively.

Simultaneously, at the opposite end of the conductivity spectrum, there has also been a large interest in two-dimensional (2D) and layered superconductors \cite{Xi,Lu,Fatemi,Sajadi,Cao,Park,FeSe,Klemmbook,Klemmpristine}.  Particular interest has been in monolayer FeSe, in  monolayer and few-layer samples of the transition metal dichalcogenide  2$H$-NbSe$_2$, in gated bulk samples of the transition metal dichalcogenide 2$H$-MoS$_2$, which also resulted in effective monolayer superconductors \cite{Xi,Lu,Fatemi,Sajadi}, and in twisted-bilayer and twisted-trilayer graphene, which was surprisingly also shown to be superconducting for the magic twist angle $\sim1.1^{\circ}$ for certain induced carrier densities \cite{Cao,Park}.

But what about the center of the conductivity spectrum?  Some very interesting 2D examples are  the metallic phases of magic-angle, twisted-bilayer and twisted-trilayer graphene obtained either with different induced carrier densities than those used to study the superconductivity \cite{Cao,Park}, or for temperatures $T$ exceeding the superconducting transition temperature $T_c$.  More generally,  by reducing the thickness of a large variety of semiconducting or insulating transition metal dichalcogenides ($MX_2$ with $M$ = Mo, V, W, Ta and $X$ = S, Se, Te) with octahedral 1$T$ or distorted octahedral 1$T'$ structures \cite{Klemmbook}, to monolayer thicknesses, the  resulting materials, possibly except for monolayer 1$T'$-WTe$_2$, turned out to be very surprisingly metallic \cite{WangY,Wu2,ZhangZ1,ZhangZ2,Xu,Lin,Sugawara,Huan,Shivayogimath}.  Presently, there appear  to be a variety of reasons for this, among which is the removal of the charge-density waves  present in bulk materials \cite{Klemmpristine}.  In addition, monolayer  MoN has been shown to be metallic \cite{Xiao,ZhangQ}, and first-principles predictions of monolayer metallicity in  Au$_2$B and MoSi$_2$ have been presented \cite{WangZ,Huang}.  The more recent growth of such crystals on Au substrates appears to yield more uniform monolayer metals than their earlier growth on NaCl microcrystals \cite{Shivayogimath, Huan}, which suggests that the number of high-quality monolayer metallic systems could increase significantly in the near future.
 In addition, ultrathin single crystals of ultrapure Al could be effectively 2D metals suitable for experiments to measure quantum effects.
Those studies suggested that such non-magnetic, metallic monolayers might have many practical uses \cite{WangY,Wu2,ZhangZ1,ZhangZ2,Xu,Lin,Sugawara,Huan,Shivayogimath,Xiao,ZhangQ,WangZ,Huang}.

However, one could also ask if there might be some qualitatively new physics that could be extracted from 2D metals.  Here we show that the answer is yes!  The quantum spin Hall effect that does not include the spin-orbit coupling interaction can be observed and used as a tool to probe the microscopic parameters of the metal, provided that the proper geometry and experimental probes are chosen.  Moreover, it could also provide a basis for the construction of a quantum computer that could potentially operate at room temperature, as for  photon-based quantum computers presently under study\cite{CIO}.

\begin{figure}
\center{\includegraphics[width=0.49\textwidth]{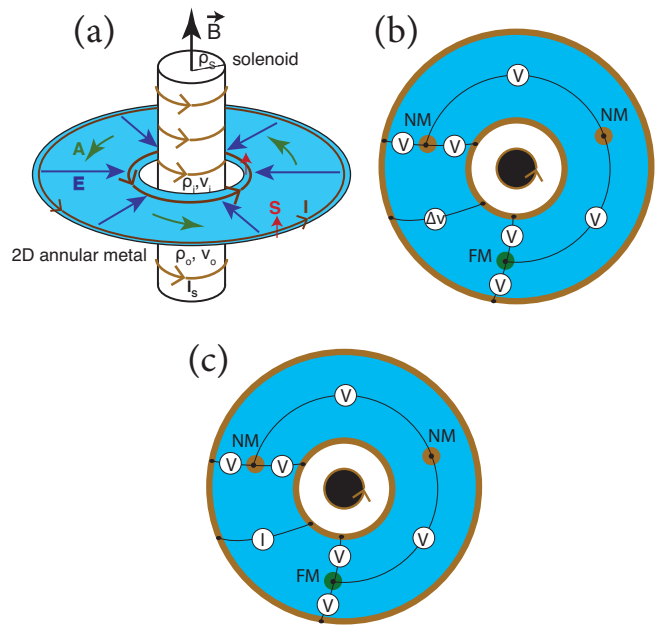}
\caption{{\bf Sketch of a side view of the non-thermally-managed  version of the proposed experimental setup}. (a). A current $I_S$ is applied to the tightly wrapped wire coil surrounding a  long cylindrical solenoid of radius $\rho_S$ that is normally placed inside a 2D metallic Corbino disk of inner radius $\rho_i > \rho_S$, generating a constant ${\bm B}_S$ inside the solenoid  and only an azimuthal vector potential ${\bm A}_S$ in the disk. (b). (top view) In the first experiment,  the electric potentials are fixed respectively at $v_i$ and $v_o$ on $\rho_i$ and $\rho_o$, respectively, inducing the radial ${\bm E}$.  (c). (top view) In the second experiment, an applied radial current ${\bm I}$ source  across $\rho_i$ and $\rho_o$ induces a uniform radial sheet charge current ${\bm K}$,  which induces the radial electric field ${\bm E}$ and the different equilibrium potentials $v_i$ and $v_o$ on $\rho_i$ and $\rho_o$.  In both cases, the combined radial ${\bm E}$ and azimuthal ${\bm A}$  couple to the electron or hole spins, leading to a spin imbalance that generates an azimuthal spin current detected by the voltage between the radially central ferromagnetic (FM) electrode and the electrically connected radially central normal metallic (NM) electrode, which should be detectably different from the charge current detected by the voltage between that radially central  NM electrode  its electrically connected radially central NM electrode.   The radial voltages between both the radially central FM and NM electrodes and either the electrodes at $\rho_i$ or $\rho_o$ should detect the same radial charge current, but no radial spin current. See text. }}
\end{figure}

\section{The proposed experiments}
We propose two different quantum spin Hall experiments that employ the previously untested but relativistically-generated quantum spin Hall Hamiltonian that does not make use of its spin-orbit coupling component.  In both experiments, a  cylindrical solenoid of radius $\rho_S$ is normally placed in the center of a 2D metallic Corbino disk of outer and inner radii $\rho_o$ and $\rho_i>\rho_S$, as sketched in Fig. 1(a).  The experimenter applies a charge current  $I_S$ in a wire tightly wrapped around the solenoid, which generates an azimuthal magnetic vector potential
\begin{eqnarray}
{\bm A}_S(\rho)&=&\hat{\varphi}\Phi_S/(2\pi\rho)\label{A}
\end{eqnarray}
in the  disk, where ${\bm B}_S={\bm\nabla}\times{\bm A}_S$,  $\Phi_S=\pi|{\bm B}_S|\rho_S^2$ is the magnetic flux in the solenoid, and due to the Biot-Savart law, $\Phi_S=\mu_0I_S\rho_S/2$, where $\mu_0$ is the magnetic permeability of vacuum.   The solenoid must be long enough that ${\bm B}_S\approx0$ everywhere in the disk, as in the  Aharonov-Bohm experiment \cite{AB,Chambers}.
\subsection{The first experiment}
In the first experiment, a uniform potential difference $\Delta v=v_o-v_i$ is applied across the outer and inner disk radii $\rho_o$ and $\rho_i$, respectively, as pictured in Fig. 1(b).  The circular electrodes must be much better conductors than the 2D metal in the Corbino disk.  This has the result of imposing the electric potentials $v_o$ and $v_i$ on $\rho_o$ and $\rho_i$, respectively.

 Since the electric potential $\Phi(\rho)$ in the disk satisfies the Laplace equation, ${\bm\nabla}^2\Phi=0$, its solution in polar coordinates is easily found to be
\begin{eqnarray}
\Phi_1(\rho)&=&v_i+\overline{\Delta v}_1\ln(\rho/\rho_i),\label{Phi}\\
\overline{\Delta v}_1&=&(v_o-v_i)/\ln(\rho_o/\rho_i),\label{Delta v}
\end{eqnarray}
so that
\begin{eqnarray}
{\bm E}_1=-{\bm\nabla}\Phi(\rho)&=&-\hat{\rho}\overline{\Delta v}_1/\rho\label{E}.
\end{eqnarray}
The combined generations of the azimuthal ${\bm A}_S$ and radial ${\bm E}_1$ given respectively by Eqs. (\ref{A}) and (\ref{E}) are employed to perform the quantum spin Hall measurements.  For this experiment, ${\bm E}_1={\bm j}_1/\sigma$ generates a radial sheet current ${\bm K}_1=K_1\hat{\bm\rho}={\bm j}_1/(2\pi\rho)$, which also generates a radial magnetic vector potential ${\bm A}_1$ surrounding the disk.  The exact wave functions and energies for independent electrons or holes in the disk are found and are presented in Section III.
\subsection{The second experiment}
In the second experiment pictured in Fig. 1(c), the potential difference $\Delta v$ pictured in Fig. 1(b) is replaced by an applied radial current ${\bm I}$, resulting in a uniform radial sheet current ${\bm K}_2$ across the disk, that also generates a radial magnetic vector potential ${\bm A}_2$ surrounding the disk.  In this case, after thermal equilibrium is attained, the applied radial sheet current ${\bm K}_2=K_2\hat{\bm\rho}$ yields a current density ${\bm j}_2={\bm K}_2/(2\pi\rho)$ and due to Ohm's law, a time-independent radial electric field  ${\bm E}_2={\bm j}_2/\sigma={\bm K}_2/(2\pi\rho\sigma)$, where $\sigma$ is the electrical conductivity  of the metallic disk. As for the first experiment, Eqs. (\ref{Phi})-(\ref{E}) apply, but in this case $\overline{\Delta v}_2=-K_2/(2\pi\sigma)$, so that Eq. (2) is replaced by
\begin{eqnarray}
\Phi_2(\rho)&=&v_i+\overline{\Delta v}_2\ln(\rho/\rho_i)=v_i-\frac{K_2}{2\pi\sigma}\ln(\rho/\rho_i),\label{Phi2}
\end{eqnarray}
where we have chosen the arbitrary integration constant to be $v_i$ in order that the two potentials may be written as
\begin{eqnarray}
\Phi_{\ell}(\rho)&=&v_i+\overline{\Delta v}_{\ell}\ln(\rho/\rho_i)\label{Phil}
\end{eqnarray}
for $\ell=1,2$. For both $\ell=1,2$, ${\bm\nabla}\cdot{\bm E}_{\ell}=({\bm\nabla}\times{\bm E}_{\ell})\cdot{\bm\sigma}=0$.

  In both experiments, the radial sheet current ${\bm K}_{\ell}$ also generates a radial vector potential ${\bm A}_{\ell}$ for $\ell=1,2$, due to Amp{\`e}re's law, as discussed in Subsection C, which contributes to the wave functions and to the radial and azimuthal charge currents, but not to either the radial or the azimuthal spin current, as discussed along with the exact wave functions in Sec. III. Finite temperature effects are discussed in Section IV.  In Section V, we summarize our results and discuss how multiple Corbino disks surrounding multiple solenoids could provide a novel technique to construct a quantum computer that could operate at temperatures well above the metal's superconducting transition and, if properly thermally managed, potentially even at room temperature.

\subsection{The induced radial vector potential in the two experiments}
For the cases $\ell=1,2$  pictured in Figs. 1(b) and  1(c), respectively, in which either a uniform potential difference or a radial current is applied across $\rho_i$ and $\rho_o$, there is an additional complication due to the induced radial vector potential given in general coordinates by \cite{Jackson}
\begin{eqnarray}
{\bm A}_{\ell}({\bm x})&=&\frac{\mu_0}{4\pi}\int\frac{{\bm j}_{\ell}({\bm x}')d^3{\bm x}'}{|{\bm x}-{\bm x}'|},
\end{eqnarray}
where $\mu_0$ is the magnetic permeability of vacuum.
Since in cylindrical coordinates, we have
\begin{eqnarray}
{\bm j}_{\ell}({\bm x}')&=&\frac{{\bm K}_{\ell}}{2\pi\rho'}\delta_{z',0}\Theta(\rho'-\rho_i)\Theta(\rho_o-\rho'),
\end{eqnarray}
where $\Theta(x)$ is the Heaviside step function,
we obtain the general expression for a long solenoid
\begin{eqnarray}
{\bm A}_{\ell}(\rho,z)&=&\frac{\mu_0{\bm K}_{\ell}}{2\pi^2}\int_{\rho_i}^{\rho_o}\frac{K(k)d\rho'}{\sqrt{\rho^2+(\rho')^2+z^2}},\label{Arho}\\
k^2&=&\frac{2\rho\rho'}{\rho^2+(\rho')^2+z^2},
\end{eqnarray}
where $K(k)$ is the elliptic integral of the first kind, which should not be confused with the sheet currents ${\bm K}_{\ell}=K_{\ell}\hat{\bm\rho}$ for $\ell=1,2$.    It is most likely easier experimentally to apply the current across those radii, $\ell=2$, but an experimental measure of this  correction could be made by doing the potential difference experiment, $\ell=1$,  at least once.
In Fig. 2, we plot $\overline{A}(\rho/\rho_i,z/\rho_i)=2\pi^2A_{\ell}(\rho/\rho_i,z/\rho_i)/(\mu_0K_{\ell})$ for  $\rho_o/\rho_i=10$ as a function of $\rho/\rho_i$ from 1 to 20 for the indicated values of $z/\rho_i$.

\begin{figure}
\center{\includegraphics[width=0.49\textwidth]{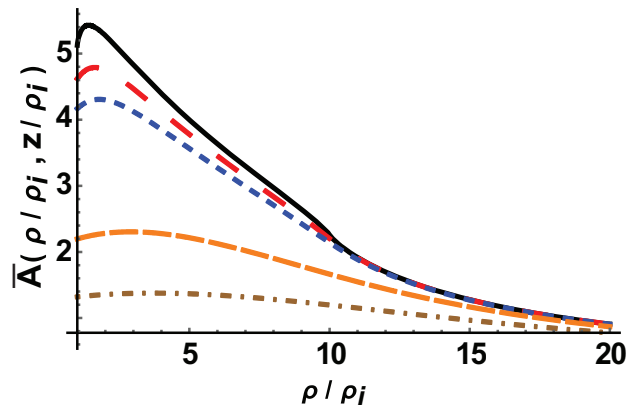}
\caption{{\bf Magnetic vector potential generated by the radial applied current or potential.} Plots of the dimensionless $\overline{A}(\rho/\rho_i,z/\rho_i)=2\pi^2A_{\ell}/(\mu_0K_{\ell})$ for $\rho_o/\rho_i=10$ in each curve, and for $z/\rho_i=10^{-4}$ (solid black), 0.5 (dashed red), 1.0 (dashed blue), 5.0 (dashed orange), and 10.0 (brown dot-dashed).   }}\label{fig2}
\end{figure}

In non-magnetic metals, it is now well established that a spin current can be generated into the metal by injecting a charge current from a ferromagnetic (FM) electrode, and measuring the voltage between two  different FM electrodes \cite{Silsbee1985,Chen2013,Tang2014,Han2020}.  In the present proposed experiments, the quantized spin current in the non-magnetic Corbino disk is spontaneously generated by  the radial electric field ${\bm E}_i$ generated either by the applied  $\Delta v$ or by the applied radial  ${\bm I}$ and by the azimuthal ${\bm A}_S$ generated by the applied current $I_S$ in the solenoid, and their combined quantum  spin Hall coupling of ${\bm E}_i$ and ${\bm p}-q{\bm A}_S$, where $q=\mp|e|$ for electrons or holes, respectively,  to the metallic electron or hole spins.

As shown in Section III, the nonvanishing  quantum spin Hall interaction in 2D,
which contains the 2D gauge-invariant operator ${\bm E}_i\times({\bm p}-q{\bm A}_S)\cdot{\bm\sigma}$, which spontaneously generates azimuthal charge and spin current densities, but does not generate any radial currents.  To measure the generated currents, two normal metallic (NM) electrodes and one FM electrode are as azimuthally equally   spaced as experimentally possible at $\rho_{\rm expt}$, also chosen to be as close to the midpoints between $\rho_i$ and $\rho_o$ as experimentally possible, as sketched in Figs. 1(b) and 1(c).  To measure the spontaneously generated  azimuthal charge and spin current densities $j^{c,s}_{\varphi}(\rho,\varphi)$, the experimenter measures the voltage $V$ between the two neighboring radially central NM electrodes and between the FM electrode and its neighboring radially central NM electrode, respectively, all at $\rho_{\rm expt}$.  To measure the total radial charge and vanishingly small spin current densities $j^{c,s}_{\rho}(\rho,\varphi)$, the experimenter respectively measures the voltage across either the NM or FM electrodes and the circular NM electrode on either $\rho_i$ or $\rho_o$, for both signs of $I$ or $\Delta v$, as pictured in Figs. 1(b) and 1(c).  Performing both of these measurements allows the experimenter to correct for slight variations in the values of $\rho_{\rm expt}$ for those two electrodes.

In Section III, we present the quantum Hamiltonian for the two  experiments, demonstrate that the 2D particle current is satisfied, treat in indeced vector potential ${\bm A}_{\ell}$ by a gauge transformation, and, except for a small potential perturbation, solve for the exact wave functions and charge and spin currents by the standard separation of radial and azimuthal variables technique. In Section IV, we discuss finite temperature effects, provide a possible technique of thermal management to allow for the experiments' operations at room temperature, and suggest a design of  thermally-managed and electrically connected, stacked Corbino disks surrounding an array of solenoids, all of which could function as a quantum computer.   In Section V,  we summarize our results.

\section{Theory of the experiments}
\subsection{The Hamiltonian and particle continuity}

For the experiment pictured in Fig. 1(a) and either  1(b) or 1(c), the Hamiltonian for a single electron or hole in a metallic 2D Corbino disk  with an isotropic planar effective mass $m_{||}$ is
\begin{eqnarray}
 H_{2D}^{(\ell)}=&H^{T,(\ell)}_{2D}+q\Phi_{\ell}(\rho)+H^{QSH,(\ell)}_{2D}\label{H2D},
 \end{eqnarray}
where $\ell=1,2$,
\begin{eqnarray}
H_{2D}^{T,(\ell)}&=&\frac{\beta[{\bm p}-q({\bm A}_S+{\bm A}_{\ell})]^2}{2m_{||}} \label{HT2D}
\end{eqnarray}
is the gauge-invariant kinetic energy,  ${\bm p}=-i\hbar{\bm\nabla}$ is the quantum-mechanical momentum in polar coordinates, ${\bm p}-q({\bm A}_S+{\bm A}_{\ell})$ is the total mechanical momentum \cite{GS,SN}, $\beta=\pm1$  for electrons or holes, respectively, $q\Phi_{\ell}(\rho)$ is the potential energy for both experiments, where ${\bm A}_S$ is given by Eq. (\ref{A}), the $\Phi_{\ell}(\rho)$ are given by Eqs. (\ref{Phi}) and (\ref{Phi2}),
and the $H^{QSH ,(\ell)}_{2D}$ are given by
\begin{eqnarray}
H^{QSH,(\ell)}_{2D}&=&-\frac{\mu_{B}}{2m_{||}c^2}[{\bm E}_{\ell}\times({\bm p}-q{\bm A}_S)]\cdot{\bm\sigma},\label{HQSH2D}
\end{eqnarray}
which
is the quantum-mechanical gauge-invariant quantum spin Hall Hamiltonian in 2D \cite{ZGHK,ZZGK}, where $\mu_B=q\hbar/(2{\rm m})$ is $\mp$ the Bohr magneton for electrons or holes, respectively, ${\rm m}$ is the rest mass of a free electron, $c$ is the speed of light in vacuum,  $\hbar=h/(2\pi)$, and the components of ${\bm\sigma}$ are the Pauli matrices representing the spin states of an electron or hole. With ${\bm A}_S$ and ${\bm E}_{\ell}$ respectively in the azimuthal and radial directions of a 2D Corbino disk, the only relevant Pauli matrix is $\sigma_z$, and the generated ${\bm A}_{\ell}$ do not enter this quantum spin Hall Hamiltonian. For the particular 2D metal under study, $m_{||}$ can be measured by cyclotron resonance with an applied ${\bm B}$ normal to the film.  Particle-particle interactions are neglected.

The density operators $\tilde{\rho}_{\ell}=\sum_s|\Psi^{\ell}_s(\rho,\varphi)|^2$ for the $\ell=1,2$ experiments satisfy \cite{Gottfried}
\begin{eqnarray}
\frac{\partial\tilde{\rho}_{\ell}}{\partial t}&=&\frac{1}{i\hbar}\sum_s\Bigl[\Psi^{\ell (*)}_sH_{2D}^{(\ell)}\Psi^{\ell}_s-\Bigl(H_{2D}^{(\ell)}\Psi^{\ell}_s\Bigr)^{\dagger}\Psi^{\ell}_s\Bigr],
\end{eqnarray}
leading to
the two-dimensional continuity equation
\begin{eqnarray}
\frac{\partial\tilde{\rho}_{\ell}}{\partial t}+{\bm \nabla}\cdot{\bm j}_{\ell}&=&0,
\end{eqnarray}
where the conserved particle currents ${\bm j}_{\ell}$ are given by
\begin{eqnarray}
{\bm j}_{\ell}&=&\frac{1}{m_{||}}{\cal R}\biggl(\sum_s\Psi_s^{\ell *}(\rho,\varphi)\biggl[\hat{\bm\rho}\Bigl(\frac{\hbar\partial}{i\partial\rho}-qA_{\ell}\Bigr)\nonumber\\
& &+\hat{\bm\varphi}\Bigl(1-\frac{\mu_BE_{\ell}s}{c^2}\Bigr) \Bigl(\frac{1}{\rho}\frac{\hbar\partial}{i\partial\varphi}-qA_S\Bigr)
\biggr]\Psi^{\ell}_s(\rho,\varphi)\biggr).\>\>\>\label{current}
\end{eqnarray}
We note that the conserved particle currents contain both  radial and azimuthal particle currents, and an azimuthal spin current driven solely by $H_{2D}^{QSH, (\ell)}$.

At this point,  a strong analogy with the Aharonov-Bohm experiment is evident.  The radial ${\bm A}_{\ell}=\hat{\bm\rho}A_{\ell}(\rho)$ can be removed from the $H_{2
d}^{T,(\ell)}$ in Eq. (\ref{HT2D}) and in Eq. (\ref{current}) for the conserved current by a gauge transformation,
\begin{eqnarray}
\tilde{\Psi}^{\ell}_s(\rho,\varphi)&=&\Psi^{\ell}_s(\rho,\varphi)\exp\Bigl[\frac{-iq}{\hbar}\int_{\rho_i}^{\rho}A_{\ell}(\rho')d\rho'\Bigr],\label{gauge}
\end{eqnarray}
where $\Psi_s^{\ell}$ is now the wave function without ${\bm A}_{\ell}$, and the full wave function with ${\bm A}_{\ell}$ is $\tilde{\Psi}^{\ell}_s$.  On the other hand,
the azimuthal vector potential ${\bm A}_S$ is quantized, and cannot be removed by a gauge transformation, as detailed in the following.

With the gauge transformation, the conserved particle currents satisfy
\begin{eqnarray}
{\bm j}_{\ell}&=&\frac{1}{m_{||}}{\cal R}\biggl(\sum_s\tilde{\Psi}_s^{\ell *}(\rho,\varphi)\biggl[\hat{\bm\rho}\Bigl(\frac{\hbar\partial}{i\partial\rho}\Bigr)\nonumber\\
& &+\hat{\bm\varphi}\Bigl(1-\frac{\mu_BE_{\ell}s}{c^2}\Bigr) \Bigl(\frac{1}{\rho}\frac{\hbar\partial}{i\partial\varphi}-qA_S\Bigr)
\biggr]\tilde{\Psi}^{\ell}_s(\rho,\varphi)\biggr).\>\>\>\>\label{current2}
\end{eqnarray}

Before writing the Hamiltonian for the wave functions $\tilde{\Psi}^{\ell}_s$ in polar  $(\rho,\varphi)$ coordinates, we first break up the Hamiltonian into two parts, one exactly solvable, and the other solvable in first order perturbation theory. We write
\begin{eqnarray}
H_{2D,s}^{(\ell)}&=&H_{2D,s}^{(\ell),0}+H_{2D}^{(\ell),1},
\end{eqnarray}
where
\begin{eqnarray}
H_{2D,s}^{(\ell),0}&=&qv_i
-\frac{\beta\hbar^2}{2m_{||}\rho^2}\biggl[\rho\frac{\partial}{\partial\rho}\Bigl(\rho\frac{\partial}{\partial\rho}\Bigr)+\frac{\partial^2}{\partial\varphi^2}\nonumber\\
& &-\delta\Bigl(\delta+2i\frac{\partial}{\partial\varphi}\Bigr)
+\frac{\gamma_{\ell}}{2\beta}\Bigl(\delta+i\frac{\partial}{\partial\varphi}\Bigr)s\biggr],\label{H2D}
\end{eqnarray}
where
\begin{eqnarray}
\delta&=&\frac{\Phi_S}{\Phi_0},\label{delta}\\
\gamma_{\ell}&=&\frac{q\overline{\Delta v}_{\ell}}{{\rm m}c^2},\label{gamma}
\end{eqnarray}
$\Phi_0=h/q$ is the flux quantum for a hole or minus the flux quantum for an electron, and
\begin{eqnarray}
H_{2D}^{(\ell),1}&=&q\Phi_{\ell}(\rho)-qv_i=q\overline{\Delta v}_{\ell}\ln(\rho/\rho_i)=\gamma_{\ell}{\rm m}c^2\ln(\rho/\rho_i),\label{H1}\nonumber\\
\end{eqnarray}
and since $|\gamma_{\ell}|\ll1$, this is a small perturbation for the entire range of experimental values of $q\overline{\Delta v}_{\ell}$, which is described in detail in Appendix A.

We then write the time-independent Schr{\"o}dinger equation for the bare (solvable) Hamiltonian as
\begin{eqnarray}
H_{2D,s}^{(\ell),0}\tilde{\Psi}_{s}^{\ell,0}(\rho,\varphi)&=&E_s^{(\ell),0}\tilde{\Psi}_{s}^{\ell,0}(\rho,\varphi).
\end{eqnarray}
Then multiplying $H_{2D}^{(\ell),0}\tilde{\Psi}^{\ell,0}_{s}-E^{(\ell),0}_s\tilde{\Psi}_s^{\ell,0}=0$ by $\rho^2$,  using the standard ``separation of variables'' technique to write $\tilde{\Psi}_s^{\ell,0}(\rho,\varphi)=R_{\ell,s}(\rho)\chi(\varphi)$ \cite{GS,SN},  and  dividing both sides of the equation by $R_{\ell,s}(\rho)\chi(\varphi)$, it is easily seen that
\begin{eqnarray}
F_{\ell,s}(\rho)&=&G_{\ell,s}(\varphi)+C_{\ell,s},
\end{eqnarray}
where
\begin{eqnarray}
F_{\ell,s}(\rho)&=&-\frac{\beta\hbar^2\Bigl[\rho\frac{d}{d\rho}\Bigl(\rho R_{\ell,s}'(\rho)\Bigr)\Bigr]}{2m_{||}R_{\ell,s}(\rho)}+\rho^2\Bigl(qv_i-E_s^{(\ell,0)}\Bigr),\nonumber\\
G_{\ell,s}(\varphi)&=&\frac{\beta\hbar^2}{2m_{||}}\Biggl[\frac{\chi''(\varphi)}{\chi(\varphi)}-i\frac{\chi'(\varphi)}{\chi(\varphi)}\Bigl(2\delta-\frac{\gamma_{\ell}}{2\beta}s\Bigr)\Biggr],\nonumber\\
C_{\ell,s}&=&\frac{\beta\hbar^2}{2m_{||}}\Bigl(-\delta^2+\frac{\gamma_{\ell}\delta}{2\beta}s\Bigr),\label{separation}
\end{eqnarray}
and where the primes and the double prime refer to the first and second derivatives with respect to the relevant spatial variable. We note that since $s=\pm1$ arises from the diagonal Pauli matrix $\sigma_z$, these equations are implicitly the diagonal elements of rank-2 matrices, so that terms not containing $s$ are implicitly proportional to ${\bm 1}$, the rank-2 identity matrix, so $F_{\ell,s}(\rho)$, $G_{\ell,s}(\varphi)$,  $C_{\ell,s}$, and the overall Hamiltonian are diagonal rank-2 matrices, or they could be rank two column vectors in the Nambu representation.

The fundamental assumption of the separation of variables technique is that $G_{\ell,s}(\varphi)=C_{\ell,s}'$ and $F_{\ell,s}(\rho)=C_{\ell,s}+C_{\ell,s}'$ must  both be constants, independent of either variable $\rho$ or $\varphi$ \cite{GS,SN}.  Since we must require $\chi(\varphi+2\pi)=\chi(\varphi)$ to be invariant under rotations by $2\pi$, we then may set
\begin{eqnarray}
\chi(\varphi)&=&e^{in\varphi},\label{chiofvarphi}
\end{eqnarray}
 where $n$ is an integer, which can be 0 or of either sign, and let its normalization constant  be included in $R(\rho)$.  We note that more complicated forms for $\chi(\varphi)$ satisfying the required  rotational invariance by $2\pi$ such as $\chi(\varphi)=a_1e^{in\varphi}+a_2e^{-in\varphi}$ are not allowed.  Although such a solution would satisfy $\chi''(\varphi)/\chi(\varphi)=-n^2$, a constant,  $\chi'(\varphi)/\chi(\varphi)$ for that form with both nonvanishing $a_1$ and $a_2$ would depend strongly upon $\varphi$, violating the fundamental assumption of the separation of variables technique that $G_{\ell,s}(\varphi)$ be a constant.  Hence, there is only one quantum number $n$ in the exponential expression for $\chi(\varphi)$, Eq. (\ref{chiofvarphi}).

Then, using  Eq. (\ref{chiofvarphi}) in the expression for $G_{\ell,s}(\varphi)$, and combining that result with the expression for $C_{\ell,s}$, we obtain
\begin{eqnarray}
C_{\ell,s}+C_{\ell,s}'&=&-\frac{\beta\hbar^2}{2m_{||}}(n-\delta)(n-\delta+\frac{\gamma_{\ell}}{2\beta}s).
\end{eqnarray}
For simplicity, we may set $\beta=1/\beta$, which is satisfied for both electrons and holes. Then,
by multiplying the resulting expression for $F_{\ell}(\rho,s)$ in Eq. (\ref{separation})  by $R_{\ell,s}(\rho)/\rho^2$, it is then elementary to obtain
 the time-independent radial Schr{\"o}dinger wave equation for $R_{\ell,s}(\rho)$,
 \begin{eqnarray}
\biggl( -\frac{\beta\hbar^2}{2m_{||}}\Bigl[\frac{1}{\rho}\frac{d}{d\rho}\Bigl(\rho\frac{d}{d\rho}\Bigr)\Bigr]+V^{\ell}_{{\rm eff},s}(\rho)\biggr)R_{\ell,s}(\rho)&=&E_s^{(\ell,0)}R_{\ell,s}(\rho),\>\>\>\nonumber\\
 \end{eqnarray}
 where the effective radial potential $V^{\ell}_{{\rm eff},s}(\rho)$ is given by
\begin{eqnarray}
V^{\ell}_{{\rm eff},s}(\rho)&=&\frac{\beta\hbar^2 z^{\ell}_{n,s}}{2m_{||}\rho^2}+qv_i,\label{Veff}\\
z^{\ell}_{n,s}&=&(n-\delta)(n-\delta+s\beta\gamma_{\ell}/2).\label{zns}
\end{eqnarray}
If $\delta=n$, the two spin states are degenerate.  The experimenter has a significant amount of flexibility in choosing the spin $s$ value of the dominant probed carriers.    Since
\begin{eqnarray}
\beta\gamma_{\ell}&=&\frac{-|e|\overline{\Delta v}_{\ell}}{{\rm m}c^2}\label{betagamma}
\end{eqnarray}
 for both electrons and holes, the dominant spin state just depends upon the sign of the potential difference (or the direction of ${\bm E}$) and the sign of $n-\delta$.  If $n-\delta>0$,  the lower energy state is that for $s\beta\gamma_{\ell}<0$.   In Fig. 1(a), the $s=+$ (up) spin state for either electrons or holes has lower energy than does the down spin state for $\overline{\Delta v}>0$. For $\overline{\Delta v}_{\ell}<0$, the $s=-$ (down) spin state is lower in energy for both electrons and holes.  On the other hand, if $n-\delta<0$, those interpretations hold for the opposite signs of $\overline{\Delta v}_{\ell}$.

Although this is qualitatively similar to the Zeeman interaction, here ${\bm B}=0$ and the spin states for both electrons and holes are distinguished by changing the signs of $\overline{\Delta v}_{\ell}$ and of $n-\delta$.  There is a very interesting interplay between the roles of the flux in the solenoid and the electric potential difference upon the spin states of the electrons or holes.

 More important, it is shown in the following that the only corrections to the energy are first order in each of the three terms in this perturbation as there are no perturbative corrections to the wave functions.  Hence, this Hamiltonian is exactly soluble for any finite value of  $\gamma_{\ell}$ and ${\bm A}_{\ell}$.

\subsection{The exact wave functions}

When $\overline{\Delta v}_{\ell}=0$,  $\gamma_{\ell}=0$, and $z^{\ell}_{n,s}=(n-\delta)^2$, so the energies of the two spin states are degenerate, both for electrons and holes.   However, for $\overline{\Delta v}_{\ell}\ne 0$, the combined signs of $\delta-n$ and $\gamma_{\ell}$ flip the spins of the ground state.  During the experiments, both signs of $z^{\ell}_{n,s}$ given by Eq. (\ref{zns}) are obtained by properly varying the sign and magnitude of $I_S$ and the sign of either $\Delta v$ or $I$.

Although $z^{\ell}_{n,s}$ can be of either sign, it is convenient to set
\begin{eqnarray}
z^{\ell}_{n,s}&=&(\nu^{\ell}_s)^2,
\end{eqnarray}
where
\begin{eqnarray}
\nu^{\ell}_s=\nu^{\ell}_{s,1}+i\nu^{\ell}_{s,2}.
\end{eqnarray}
That is, if $z^{\ell}_{n,s}>0$,  $\nu^{\ell}_s=\nu^{\ell}_{s,1}$ is real, and if $z^{\ell}_{n,s}<0$, $\nu^{\ell}_s=i\nu^{\ell}_{s,2}$ is pure imaginary.

Since there are two distinct spin states characterized by $z^{\ell}_{n,s}$,
 we write $H^{\ell}_{0,s}R_{\nu^{\ell}_s}(\rho)=E^{({\ell})}_{0,s}R_{\nu^{\ell}_s}(\rho)$, and after multiplying by $-2m_{||}\rho^2\beta/\hbar^2$, we have
\begin{eqnarray}
\rho^2R''_{\nu^{\ell}_s}(\rho)+\rho R'_{\nu^{\ell}_s}(\rho)-(\nu^{\ell}_s)^2R_{\nu^{\ell}_s}(\rho)&=&-\frac{2m_{||}\beta \tilde{E}^{\ell}_{0,s}\rho^2}{\hbar^2}R_{\nu^{\ell}_s}(\rho),\nonumber\\
& &
\end{eqnarray}
where $R'_{\nu^{\ell}_s}$ and $R''_{\nu^{\ell}_s}$ are the first and second derivatives of $R_{\nu^{\ell}_s}$ with respect to $\rho$, and $\tilde{E}^{\ell}_{0,s}=E^{(\ell)}_{0,s}-qv_i$.  Then, by setting
\begin{eqnarray}
\tilde{E}^{\ell}_{0,s}&=&\beta\hbar^2(k^{\ell}_s)^2/(2m_{||}),\label{E0}
\end{eqnarray}
 and letting the dimensionless variables be $x^{\ell}_s=k^{\ell}_s\rho$, we have
\begin{eqnarray}
(x^{\ell}_s)^2R''_{\nu^{\ell}_s}(x_s)+x^{\ell}_sR'_{\nu^{\ell}_s}(x^{\ell}_s)+[(x^{\ell}_s)^2-(\nu^{\ell}_s)^2]R_{\nu^{\ell}_s}(x^{\ell}_s)&=&0,\nonumber\\
& &
\end{eqnarray}
each $\nu^{\ell}_s$ values of which are Bessel equations.

 The solutions relevant to this problem are the Hankel functions $H_{\nu^{\ell}_s}^{(1)}(x^{\ell}_s)=J_{\nu^{\ell}_s}(x^{\ell}_s)+iY_{\nu^{\ell}_s}(x^{\ell}_s)$ and $H_{\nu_s}^{(2)}(x^{\ell}_s)=[H_{\nu^{\ell}_s}^{(1)}(x^{\ell}_s)]^{*}$ \cite{AS}, so the general wave functions are
\begin{eqnarray}
\Psi^{\ell}_{n,s}(\rho,\varphi)&=&[B^{\ell}_sH_{\nu^{\ell}_s}^{(1)}(k^{\ell}_s\rho)+C^{\ell}_sH_{\nu^{\ell}_s}^{(2)}(k^{\ell}_s\rho)]e^{in\varphi},\label{Rnus}
\end{eqnarray}
where $B^{\ell}_{s}$ and $C^{\ell}_{s}$ are constants that generally depend upon the boundary conditions, and in the presence of ${\bm A}_{\ell}=\hat{\bm\rho}A_{\ell}(\rho,z=0)$,  after making a gauge transformation, the full wave function including $A_{\ell}(\rho)$ is given by Eq. (17).

  For both $\ell=1,2$, ${\bm\nabla}\cdot{\bm E}_{\ell}=({\bm\nabla}\times{\bm E}_{\ell})\cdot{\bm\sigma}=0$.
 We note that $H^{(1)}_{\nu^{\ell}_s}(k^{\ell}_s\rho)$ and $H^{(2)}_{\nu^{\ell}_s}(k^{\ell}_s\rho)$ are respectively the outward and inward radial waves.

Since $\rho_i$ and $\rho_o$ are macroscopic quantities, and the $k^{\ell}_s$ are the wave vectors of the metallic 2D annulus, we expect $k^{\ell}_s\sim\frac{2N\pi}{a}$, where $N\ge1$ and  $a$ is on the order of a lattice constant, which is much less than $\rho_i$.
 Then, for all $\rho_i\le\rho\le \rho_o$, $k^{\ell}_s\rho_i\gg1$, and the large $k^{\ell}_s\rho$ asymptotic forms of the Hankel functions are valid,
\begin{eqnarray}
H_{\nu}^{(1)}(x)&\approx&\sqrt{\frac{2}{\pi x}}e^{i(x-\pi/4-\pi\nu/2)}.
\end{eqnarray}
 Then,  the asymptotic
forms of the Hankel functions that can describe either sign of $z^{\ell}_{n,s}$ are
\begin{eqnarray}
H_{\nu^{\ell}_s}^{(1)}(x^{\ell}_s)&=&[H_{\nu^{\ell}_s}^{(2)}(x^{\ell}_s)]^{*}\nonumber\\
& \approx&\sqrt{\frac{2}{\pi x^{\ell}_s}}\exp[i(x^{\ell}_s-\eta_{\nu^{\ell}_s})+\pi\nu^{\ell}_{s,2}/2],\label{asymptotic}\\
\eta_{\nu^{\ell}_s}& =&\frac{\pi}{2}\Bigl(\nu^{\ell}_{s,1}+\frac{1}{2}\Bigr).\label{etanus}
\end{eqnarray}
The experimenters first need to provide an independent measurement of $m_{||}$ and the Fermi wave vector $k_{F,0}$ when both $\Delta v_{\ell}=0$ for $\ell=1,2$, which can be respectively measured by cyclotron resonance and angle-resolved photoemission experiments on an identical  sample of the same 2D metal.
 They are then ready to perform the main experiments, which without proper thermal management of a conventional, three-dimensional device, normally would be done at low $T$, to minimize heating effects.  There are two probes to force the spontaneously generated quantized currents:  $\delta$, or the flux $\Phi_S$ in the solenoid, controlled by the current in the wire wrapped around it, and either the applied $\Delta v$ or the applied radial current $I$.

\subsection{The charge and spin currents}

For the both experiments,  the gauge-invariant radial and azimuthal charge and spin current densities are generalizations to polar coordinates of the gauge-invariant one-dimensional particle current density \cite{GS,SN},
\begin{eqnarray}
j_{\rho}^{(\ell),c}(\rho,\varphi)&=&\frac{\hbar q}{m_{||}}{\rm Im}\Bigl[\sum_{s= \pm}\tilde{\Psi}^{(\ell)*}_{n,s}(\rho,\varphi)\frac{\partial}{\partial\rho}\tilde{\Psi}^{(\ell)}_{n,s}(\rho,\varphi)\Bigr],\\
j_{\rho}^{(\ell),s}(\rho,\varphi)&=&\frac{\hbar}{m_{||}}{\rm Im}\Bigl[\sum_{s = \pm}s\tilde{\Psi}^{(\ell)*}_{n,s}(\rho,\varphi)\frac{\partial}{\partial\rho}\tilde{\Psi}^{(\ell)}_{n,s}(\rho,\varphi)\Bigr]\nonumber\\
&=&0,\\
j^{(\ell),c}_{\varphi}(\rho,\varphi)&=&\frac{\hbar q}{m_{||}\rho}{\rm Im}\Bigl[\sum_{s = \pm}\tilde{\Psi}^{(\ell)*}_{n,s}(\rho,\varphi)\Bigl(\frac{\partial}{\partial\varphi}\Bigr)\tilde{\Psi}^{(\ell)}_{n,s}(\rho,\varphi)\Bigr]\nonumber\\
& &-\frac{q\Phi_S}{m_{||}2\pi\rho}\sum_{s=\pm}|\tilde{\Psi}^{(\ell)}_{n,s}(\rho,\varphi)|^2,
\end{eqnarray}
and
\begin{eqnarray}
j^s_{\varphi}(\rho,\varphi)&=&\frac{-\mu_BE_{\ell}\hbar}{m_{||}c^2\rho}{\rm Im}\Bigl[\sum_{s = \pm}s\tilde{\Psi}^{(\ell)*}_{n,s}(\rho,\varphi)\frac{\partial}{\partial\varphi}\tilde{\Psi}^{(\ell)}_{n,s}(\rho,\varphi)\Bigr]\nonumber\\
& &+\frac{q\Phi_S\mu_BE_{\ell}}{m_{||}c^22\pi\rho}\sum_{s=\pm}s|\tilde{\Psi}^{(\ell)}_{n,s}(\rho,\varphi)|^2,
\end{eqnarray}
where $\tilde{\Psi}^{(\ell)}_{n,s}(\rho,\varphi)$ is given by Eq. (\ref{gauge}).

\begin{figure}
\center{\includegraphics[width=0.49\textwidth]{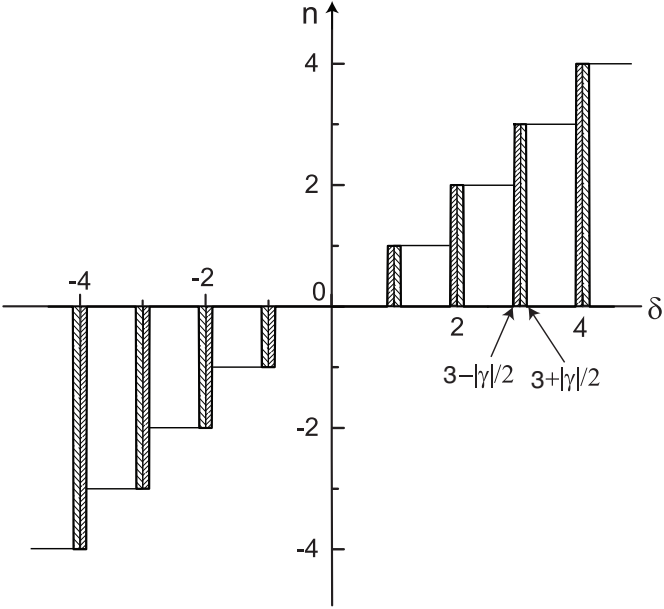}
\caption{{\bf Measurements of the quantum spin Hall effect.}  By adjusting $\delta$ and $\gamma$, the experimenters should focus on the endpoints of the regions of $\delta$ satisfying $n-|\gamma|/2\le\delta\le n+|\gamma|/2$, where the onset of the increased quantization of the azimuthal current occurs.  Note that $|\beta\gamma|=|\gamma|$. The two endpoints with $n=3$ are indicated by arrows.}}
\end{figure}

After setting  $\Delta v_{\ell}=0$ for $\ell=1, 2$, so that $\gamma=0$, the experimenters should first measure the quantum Hall effect by varying $\delta$ until it is an integer $n$, for which the spontaneous jump in the azimuthal current will occur, as sketched at the centers of the hatched regions of Fig. 3.  Then, depending upon the sign of the carrier charge $q$, they should either slightly decrease or increase $\delta$ from $n$, and increase or decrease $I$ or $\Delta v$ and hence $\gamma$. Then, due to the quantum spin Hall effect, the spontaneous jump in the azimuthal current will occur at  $\delta=n+s\beta\gamma/2$.  Depending upon which $s$ value corresponds to $E_F$, this will either be at $\delta=n+\beta\gamma/2$ or at $\delta=n-\beta\gamma/2$.  By changing the sign of $\gamma$ and repeating the experiment, the jump will switch to the other possibility. These points correspond to the dark vertical lines in Fig. 3. In that figure,  the case $n=3$ is indicated by arrows.

\section{Finite temperature effects}

\subsection{The chemical potential}

The single particle states of the non-interacting electron or hole gas  in the 2D Corbino disk are then given by
\begin{eqnarray}
\epsilon_{\ell}(k_{\ell})=\sum_s[E^{\ell}_0(k^{\ell}_s)+E^{\ell}_1(k^{\ell}_s)],
\end{eqnarray}
where the $E^{\ell}_1(k^{\ell}_s)$ are given in Appendix A,
and their states are occupied according to the Fermi-Dirac distribution function
\begin{eqnarray}
f[\epsilon_{\ell}(k_{\ell})]&=&\frac{1}{e^{[\epsilon_{\ell}(k_{\ell})-\mu_{\ell}(T)]/(k_BT)}+1},
\end{eqnarray}
where $k_B$ is Boltzmann's constant,  and for a free-particle $\epsilon_{\ell}(k_{\ell})$ in 2D, an excellent approximation to the present model, the chemical potential $\mu_{\ell}(T)$  is given by \cite{Mahan}
\begin{eqnarray}
\mu_{\ell}(T)&=&\mu_{\ell}(0)+k_BT\ln[1-e^{-\mu_{\ell}(0)/(k_BT)}]\nonumber\\
&\approx&\mu_{\ell}(0)-k_BTe^{-\mu_{\ell}(0)/(k_BT)},\label{muofT}
\end{eqnarray}
which is nearly independent of $T$, so that
\begin{eqnarray}
\mu_{\ell}(0)&\simeq&E^{\ell}_{F},\label{mu0}
\end{eqnarray}
where
\begin{eqnarray}
E^{\ell}_F&=&{\rm max}_s[E^{\ell}_0(k^{\ell}_{F,s})+E^{\ell}_1(k^{\ell}_{F,s})]\label{EF}
\end{eqnarray} is the Fermi energy,  the ground state energy of the 2D metallic Corbino disk. Equation (\ref{EF}) applies for both electrons and holes.
However, Eq. (\ref{mu0}) also implies that the $k^{\ell}_{F,s}$ have additional dependencies upon $v_i$ and $v_o$, as well as upon $\rho_i$, $\rho_o$, and the $v_0-v_i$  dependence of $\nu_s$ in $E^{\ell}_0(k^{\ell}_{F,s})$. Since for most 2D metals, $k_BT<E^{\ell}_{F}/3$ would be in the low-$T$ regime,  one could in principle  perform the experiment at room temperature with an appropriate metal.

\subsection{Thermal management}

However, with large applied currents and the resulting radial voltage difference across the disk, Joule heating could be a problem, unless the experimenters found a way to significantly reduce it.  Such heat removal is now standard with the high-transition temperature $T_c$ superconducting Bi$_2$Sr$_2$CaCu$_2$O$_{8+\delta}$ (Bi2212) terahertz emitters by coating the top and bottom of each device with Au \cite{KK1,KK2}, and sandwiching the emitter between sapphire plates with properly placed gold electrodes  \cite{KashiwagiPRApplied}, allowing it to operate in liquid $N_2$ \cite{Minami2015}. Although the metallic disks cannot be coated with Au except in the specific NM electrode positions, a similar design could work to control the Joule heating in this experiment.  The thermally-managed solenoid can be constructed by tightly wrapping a very fine electrically insulated wire coil around a cylindrical sapphire rod, which is tightly covered with a thin cylindrical sapphire sheath. The sapphire sheath could be produced by sputtering in a cooled environment.  Then,  the 2D metallic Corbino disk should be sandwiched between two identical sapphire Corbino disks, each several $\mu$m thick, several $\mu$m smaller in inner radius and similarly larger in outer  radius than the corresponding radii of the metallic disk, in order to allow them to be held together with polyimide glue near the sapphire disk radial edges and outside of the metallic disk radii \cite{KashiwagiPRApplied}. Since additional metallic holders would interfere with the experiment and screws could break the sapphire plates \cite{KashiwagiPRApplied}, gluing might be the best procedure.  The Au voltage and current electrodes  are deposited on the top of the bottom sapphire disk, which will make electrical contact with the three small electrodes on the bottom of the 2D metallic disk when the sapphire disks are tightly  glued together \cite{KashiwagiPRApplied}. The gluing should be done under a very slight pressure which is strong enough to insure the electrical contacts are all satisfactory and weak enough not to damage the 2D metallic disk.   Such a design of the thermally-managed Corbino disk is sketched in Fig. 4.
Such or similar heat control procedures could allow the  experiment to be performed at easily accessible $T$ values \cite{KashiwagiPRApplied,Minami2015}, such as in liquid $N_2$ or conceivably even up to room temperature.

\begin{figure}
\center{\includegraphics[width=0.25\textwidth]{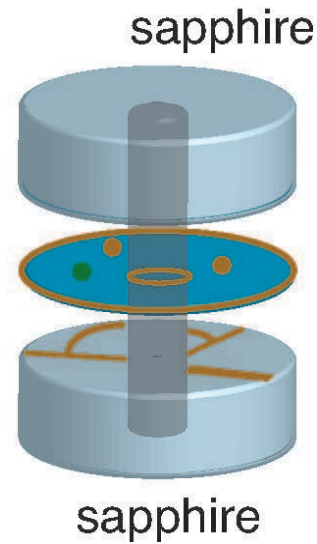}
\caption{{\bf Thermal management of the device.} The solenoid is constructed from a sapphire rod tightly wrapped in very fine electrically insulated wire tightly coated with sapphire. The 2D metallic disk is first treated with thin Au electrodes on $\rho_i$ and $\rho_o$, and then the FM and the two NM electrodes shown in Figs. 1(b) and 1(c) are pasted on one side, which in this sketch is its bottom. Then the 2D metallic Corbino disk is sandwiched between identical sapphire plates several $\mu$m in thickness, several $\mu$m smaller in inner radius than $\rho_i$ and similarly larger in outer radius than $\rho_o$. The lower sapphire plate has Au electrodes deposited on its top as sketched, which, after the three-disk sandwich is glued  together (not shown) \cite{KashiwagiPRApplied},  will become the current and voltage electrodes sketched in Figs. 1(b) and 1(c).  See text.}}
\end{figure}

 \begin{figure}
\center{\includegraphics[width=0.49\textwidth]{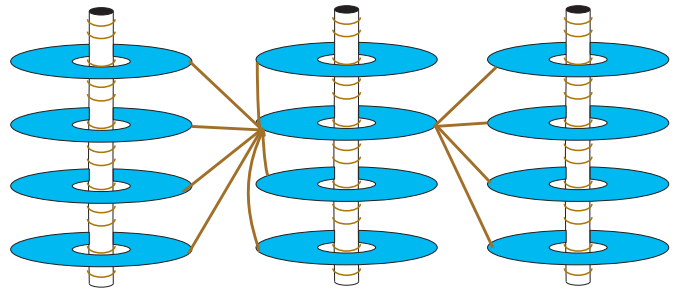}
\caption{{\bf Sketch of a  possible  quantum computer.} Shown is a sketch of a small part of a large hexagonal close-packed array of thermally-managed solenoids, as described  in the text, each with multiple thermally-managed 2D metallic Corbino disks surrounding it, with electrodes as sketched in Figs. 1(b) and (1)c. In this sketch, one of the metallic disks is connected by current leads to all of the others.  In an actual quantum computer, each disk element should be connected to all of the others. \cite{KashiwagiPRApplied,Minami2015}.}}
\end{figure}

\begin{figure}
\center{\includegraphics[width=0.49\textwidth]{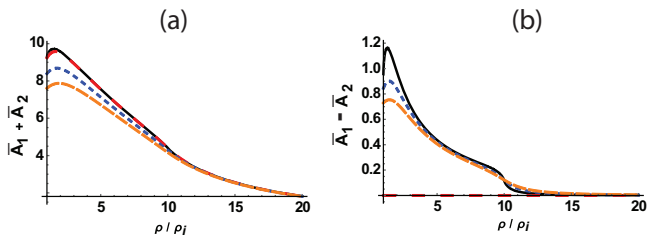}
\caption{{\bf Magnetic vector potentials  generated from Corbino  disks with the currents or potential differences in the same and opposite directions.} As in Fig. 2, we plot the dimensionless $\overline{A}=2\pi^2A_{\ell}/(\mu_0K_{\ell})$ for both disks 1 and 2 fixed at vertical positions $z/\rho_i=1.0$ and $10^{-4}$, respectively.  In (a) and (b), $\overline{A}_1+\overline{A}_2$ and $\overline{A}_1-\overline{A}_2$ are both shown for $\rho_o/\rho_i=10$ and are plotted versus $\rho/\rho_i$ for the vertical measurement positions  $z\rho_i=10^{-4}$ (solid black), 0.5 (dashed red), -0.5 (dashed blue), and -1.0 (dashed orange).}}
\end{figure}

We note that changes in the applied radial current ${\bm I}$ makes changes in $v_i$ and $v_o$ and linear changes in $E_F$ that are easy to evaluate and should be measurable.  In addition, Eq. (\ref{EFofks}) shows that the quadratic dispersion of the particles only differs for each spin by a constant, so that Eq. (\ref{muofT}) is highly accurate, implying that the experiment should be possible at room temperature, even when including the perturbations to the ground state energies.

The experimenters should first make independent measurements of the Fermi wave vector $k_{F,0}$ in the absence of $v_i$ and $v_o$ and  of $m_{||}$ (such as by angle-resolved photoemission measurements and cyclotron resonance experiments) for the 2D metal under study.  Then they should perform four current  measurements.  They should measure $j^c_{\varphi}(\rho_{\rm expt})$ and $j^s_{\varphi}(\rho_{\rm expt})$ from the two appropriate voltage leads pictured in Figs. 1(b) and 1(c), and they should also measure  $j^c_{\rho}$ and check to see that $j^s_{\rho}$ is sufficiently small, from the NM and FM electrodes at $\rho_{\rm expt}$ (roughly midway between $\rho_i$ and $\rho_o$)  and the NM electrodes at $\rho_i$ and at $\rho_o$.  As mentioned previously, they should do each of these measurements both for positive and negative $z^{\ell}_{n,s}$, which respectively result in non-vanishing $\nu^{\ell}_{s,1}$ and $\nu^{\ell}_{s,2}$.  From these measurements, experimental values for all of the wave function parameters can be determined.

 By changing $\delta$ and $\gamma$ in $\nu_s$, the experimenters can distinguish the special cases $\nu_s=0$, which can occur in three ways:  either $\delta=n$, $\delta+\beta\gamma/2=n$, or $\delta-\beta\gamma/2=n$. Since the latter two cases apply simultaneously for opposite spins of the electrons or holes, it is actually rather easy to do the experiment. Since $\gamma \ll 1$, the experimenters should choose $\delta$ very near to an integer, and then the experiment will be very sensitive to $\gamma$ variations.  For the special region for which one fraction of the electrons or holes satisfies $z^{\ell}_{n,\pm}<0$, $\nu^{\ell}_{s,1}=0$ and $\nu^{\ell}_{s,1}\ne0$, so that  $\eta^{\ell}_{\nu_s}=\pi/4+\pi\nu^{\ell}_{s,1}/2$, and the spontaneous azimuthal currents from  the particles with different spins will have different phases.  These different phases will be present simultaneously, and can be probed by varying $\Delta v_{\ell}$.   It is possible to measure the overall phase shift  $\zeta_s$ between the outgoing and incoming waves, which should be the same at each quantum jump in $j_{\varphi}$.

We note that from Eq. (\ref{Arho}), the magnetic induction ${\bm B}(\rho,z)$ generated from ${\bm A}_{\ell}$ is in the azimuthal direction, and is odd in $z$, so that it vanishes in the disk.  From the standpoint of the stack of disks evenly spaced about a single solenoid, as sketched in Fig. 5, for the applied currents or potential differences in adjacent disks equal to each other, the combined radial ${\bm A}_{\ell}$  will add,  but for the applied currents or potential difference in adjacent disk equal in magnitude and opposite in sign, the combined radial ${\bm A}_{\ell}$  will nearly cancel, as shown in Fig. 6.  This could play an important role in the entanglement issues with regard to the construction of a quantum computer.

 \section{Summary and conclusions}

 In conclusion, an experiment is proposed to measure the quantum spin Hall effect in 2D  metallic films that is qualitatively different from the effect in topological insulators, as it makes use of ${\bm E}$ and ${\bm A}$ but not of spin-orbit coupling.  The apparatus is a thermally-managed 2D metallic Corbino disk surrounding a thermally-managed cylindrical solenoid, the applied current around which generates an azimuthal ${\bm A}$ in the disk, and by applying either a uniform radial current $I$ or a potential difference $\Delta v$ between the inner and outer radii $\rho_i$ and $\rho_o$, both a radial electric field ${\bm E}$  and  fixed potentials $v_i$ and $v_o$ on those respective radii are generated.   Then, by varying the flux in the solenoid and the potentials on the inner and outer radii, quantized azimuthal charge and spin  currents are spontaneously generated.  This quantized azimuthal current can be studied by  adjusting the solenoid flux to be either an integral number of flux quanta or slightly different from an integral number of flux quanta combined with either positive or negative values of the applied radial current $I$ that couples to the carrier spins by the quantum spin Hall interaction.  No spin-orbit coupling is involved in the experiment. The relevant quantum Hamiltonian for this system is exactly soluble. Provided that the thermal management  design sketched in Fig. 4, or a modified version of it, functions as desired, the experiment could in principle be performed at room temperature. Considering an individual Corbino disk surrounding a solenoid to be a qubit, an hexagonally close-packed 2D array of thermally-managed solenoids,  each with stacks of thermally-managed 2D metallic Corbino disks surrounding it, as sketched in Fig. 5, could then function as  a  quantum computer.  However, to properly control the required entanglement of the quantum states of each qubit, it is likely that the quantum computer would operate best at a low temperature, but the temperature of liguid N$_2$ might be sufficiently low.

\section{Appendix A}
We remark that  $H^{(\ell),1}_{2D}$ given by Eq. (\ref{H1}) can be treated in first-order perturbation theory, and that higher order corrections to the wave function contain the  expectation values $\langle \tilde{\Psi}^{\ell}_{n,s}|H^{(\ell),1}_{2D}|\tilde{\Psi}^{\ell}_{n',s'}\rangle$ for $n'\ne n$ and/or $s'\ne s$.  However, we note that $\delta$ and $\beta\gamma$ in $z^{\ell}{n,s}$ are experimental parameters, so the only quantized objects are $n$, $n'$, $s$, and $s'$. But since $H^{(\ell),1}_{2D}$ is independent of $\varphi$ and of the spin $s$, we must have $n'=n$ and $s'=s$, so all higher-order corrections to the wave function vanish, as do the second and all higher-order corrections to the energy.  Hence, the exact energies $E^{\ell}_s=E^{\ell}_0(k^{\ell}_s)+E^{\ell}_1(k^{\ell}_s)$, where the $E^{\ell}_0(k^{\ell}_s)$ are given by Eq. (\ref{E0}) and

\begin{eqnarray}
E^{\ell}_1(k^{\ell}_{s})&=&\langle \tilde{\Psi}^{\ell}_{n,s}|H^{(\ell),1}_{2D}|\tilde{\Psi}^{\ell}_{n,s}\rangle\label{E1ofks}\\
&=&\frac{1}{\overline{n^{\ell}_s}}\int_0^{2\pi}d\varphi\int_{\rho_i}^{\rho_o}\rho d\rho\tilde{\Psi}^{\ell*}_{n,s}(\rho,\varphi)H_{2D}^{(\ell),1}(\rho)\tilde{\Psi}^{\ell}_{n,s}(\rho,\varphi)\nonumber\\
&=&q\overline{\Delta v}_{\ell}\ln(\rho_o/\rho_i)g^{\ell}_s(k^{\ell}_s),\nonumber\\
g^{\ell}_s(k^{\ell}_s)&=&\frac{4\rho_i}{\overline{n^{\ell}_s}\ln(\rho_o/\rho_i)}\frac{e^{\pi\nu^{\ell}_{s,2}}}{k^{\ell}_{s}}\int_1^{\rho_o/\rho_i}dx\ln(x)\Bigl[|B^{\ell}_s|^2\nonumber\\
& &+|C^{\ell}_s|^2+2{\rm Re}\Bigl(B^{\ell}_sC^{\ell*}_se^{2i(k^{\ell}_{s}\rho_ix-\eta_{\nu^{\ell}_s})}\Bigr)\Bigr],\label{gs}
\end{eqnarray}
where we have used the standard Dirac notation for the expectation value \cite{GS,SN}, the wave functions $\tilde{\Psi}^{\ell}_{n,s}(\rho,\varphi)$ are given by Eqs. (\ref{gauge}) and (\ref{Rnus})  with the asymptotic forms for the Hankel functions, Eq. (\ref{asymptotic}), $H^{(\ell),1}_{2D}(\rho)$ is given by Eq. (\ref{H1}), $\overline{n^{\ell}_s}$ is the number of particles of spin $s$ given by Eq. (\ref{overlinens}), $\eta_{\nu^{\ell}_s}$ is given by Eq. (\ref{etanus}), and we have renormalized the integration variable to the dimensionless $x=\rho/\rho_i$.  We note that $g^{\ell}_s$  could easily be integrated by parts to obtain exact solutions in terms of the sine and cosine integral functions.   But, Eq. (\ref{gs}) is already in a closed form for the exact perturbation energy $E^{\ell}_1(k^{\ell}_{s})$ for $k^{\ell}_s\rho_i\gg1$.

\begin{figure}
\center{\includegraphics[width=0.49\textwidth]{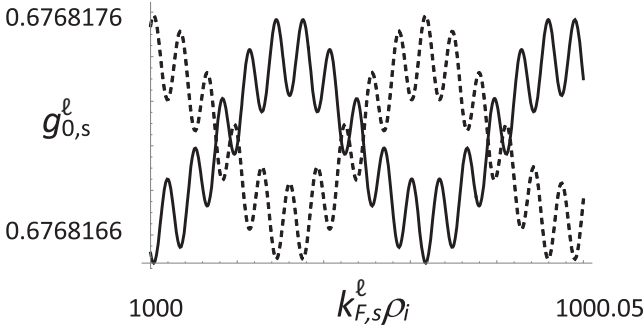}}
\caption{{\bf Plots of $g^{\ell}_{0,s}(k^{\ell}_{F,s}\rho_i,\rho_o/\rho_i,\eta'_{\nu^{\ell}_s})$ with vanishingly small radial current.} The solid and dashed curves are plots of $g^{\ell}_{0,s}(k^{\ell}_F\rho_i,10,\eta'_{\nu^{\ell}_s})$ given by Eq. (\ref{g0}) for $k^{\ell}_{F,s}\rho_i$ from 1000 to 1000.05 and $\eta'_{\nu^{\ell}_s}=\pi/2$ (solid) and $3\pi/2$ (dashed), respectively.}
\end{figure}

We now consider a special case of this perturbation correction.  For the general case, we have from
\begin{eqnarray}
E^{\ell}_F&=&{\rm max}_s\Bigl(\frac{\beta\hbar^2(k^{\ell}_{F,s})^2}{2m_{||}}+qv_i+q\overline{\Delta v}_{\ell}\ln(\rho_o/\rho_i)g^{\ell}_s(k^{\ell}_{F,s})\Bigr),\nonumber\\
\end{eqnarray}
where $g^{\ell}_s(k^{\ell}_{F,s})$ is given by Eq. (\ref{gs}) with $k^{\ell}_s\rightarrow k^{\ell}_{F,s}$. Since the  radial spin current density  given by Eq. ({\ref{spincurrent}) vanishes,
\begin{eqnarray}
C^{\ell}_s=B^{\ell}_se^{i\chi^{\ell}_s}.
\end{eqnarray}
Note that this does not mean that the quantum-induced azimuthal charge and spin currents vanish, as they are given by Eqs. (\ref{jcazimuthal}) and (\ref{jsazimuthal}).
In this particular  case, at low $T$ where one spin state dominates the other near to $E^{\ell}_F$, $g^{\ell}_s(k^{\ell}{F,s})\approx \ln(\rho_o/\rho_i)g^{\ell}_{0,s}(k^{\ell}_{F,s})$, where $g^{\ell}_{0,s}(k^{\ell}_{F,s})$ is  given by
\begin{eqnarray}
g^{\ell}_{0,s}(k^{\ell}_{F,s})\!&=&\!\frac{\int_{1}^{\rho_o/\rho_i}dx\ln(x)\cos^2(k^{\ell}_{F,s}\rho_ix-\eta'_{\nu^{\ell}_s})}{\ln(\rho_o/\rho_i)\int_{1}^{\rho_o/\rho_i}dx\cos^2(k^{\ell}_{F,s}\rho_ix-\eta'_{\nu^{\ell}_s})},\label{g0}\>\>\>\>\>
\end{eqnarray}
where $g^{\ell}_{0,s}>0$ is a dimensionless function of $k^{\ell}_{F,s}\rho_i$, $\rho_o/\rho_i$ and
\begin{eqnarray}
\eta'_{\nu^{\ell}_s}&=&\eta_{\nu^{\ell}_s}+\chi_s/2.
\end{eqnarray}
Using the identity $\cos^2z=\frac{1}{2}[1+\cos(2z)]$ in Eq. (\ref{g0}), integrating the denominator  and the non-oscillatory integral in the numerator exactly,  integrating the oscillatory term in the numerator by parts, and expanding the results in powers of $(2k^{\ell}_{F,s}\rho_i)^{-1}$, it is easy to show that
\begin{eqnarray}
g^{\ell}_{0,s}(k^{\ell}_{F,s})&=&g_{\infty}(\rho_o/\rho_i)+\frac{C_1}{2k^{\ell}_{F,s}\rho_i}+{\cal O}\Bigl(\frac{1}{2k^{\ell}_{F,s}\rho_i}\Bigr)^2,\>\>\>
\end{eqnarray}
 where
\begin{eqnarray}
{\rm lim}_{k^{\ell}_{F,s}\rho_i\rightarrow\infty}g^{\ell}_{0,s}(k^{\ell}_{F,s})&=&g_{\infty}(\rho_o/\rho_i)\nonumber\\
& =&\frac{1}{1-\rho_i/\rho_o}-\frac{1}{\ln(\rho_o/\rho_i)},\label{constant}
\end{eqnarray}
and  $C_1$ contains only terms that are oscillatory in $2k^{\ell}_{F,s}\rho_i$ and $2k^{\ell}_{F,s}\rho_o$.

In Fig. 7, plots of $g^{\ell}_{0,s}(k^{\ell}_{F,s}\rho_i,\rho_o/\rho_i,\eta'_{\nu^{\ell}_s})$ for $\rho_o/\rho_i=10$ are shown for $\eta'_{\nu^{\ell}_s}=\pi/2$ (solid) and $\eta'_{\nu^{\ell}_s}=3\pi/2$ (dashed) in the range of $k^{\ell}_{F,s}\rho_i$ from  1000 to 1000.05, respectively, in order to display the weak oscillations clearly.  We note that for $\rho_o/\rho_i=10$, the values of $g^{\ell}_{0,s}$ in these plots are in good agreement with $g_{\infty}(10)=\frac{10}{9}-1/\ln(10)\approx 0.6768166292$, as given by Eq. (\ref{constant}).  Thus, to a good approximation, when the radial induced charge and spin currents vanish, we have
\begin{eqnarray}
E^{\ell}_F&\approx&{\rm max}_s\Biggl(\frac{\beta\hbar^2(k^{\ell}_{F,s})^2}{2m_{||}}\Biggr)+qv_i\nonumber\\
& &+q\overline{\Delta v}_{\ell}\Biggl(\frac{\ln(\rho_o/\rho_i)}{1-\rho_i/\rho_o}-1\Biggr).\label{EFofks}
\end{eqnarray}

\section{Appendix B}

From the asymptotic forms of the radial wave functions, Eqs. (\ref{asymptotic}) and (\ref{etanus}),
 the radial and azimuthal charge and spin currents are independent of $\varphi$ and have the forms
 \begin{eqnarray}
j_{\rho}^c(\rho)&=&\frac{2q\hbar}{\pi m_{||}\rho}\sum_{s = \pm}e^{\pi\nu^{\ell}_{s,2}}\biggl(|B^{\ell}_s|^2-|C^{\ell}_s|^2\nonumber\\
& &-\frac{qA_{\ell}}{k^{\ell}_{F,s}}\biggl|B_s^{\ell}e^{i(k^{\ell}_{F,s}\rho-\eta_{\nu^{\ell}_s})}+
C_s^{\ell}e^{-i(k^{\ell}_{F,s}\rho-\eta_{\nu^{\ell}_s})}\biggr|^2\biggr),\label{chargecurrent}\nonumber\\
& &\\
j_{\rho}^s(\rho)&=&\frac{2\hbar}{\pi m_{||}\rho}\sum_{s = \pm}se^{\pi\nu^{\ell}_{s,2}}\biggl(|B^{\ell}_s|^2-|C^{\ell}_s|^2\nonumber\\
& &-\frac{qA_{\ell}}{k^{\ell}_{F,s}}\biggl|B_s^{\ell}e^{i(k^{\ell}_{F,s}\rho-\eta_{\nu^{\ell}_s})}+
C_s^{\ell}e^{-i(k^{\ell}_{F,s}\rho-\eta_{\nu^{\ell}_s})}\biggr|^2\biggr)\nonumber\\
&=&0,\label{spincurrent}\\
j_{\varphi}^c(\rho)&=&\frac{2(n-\delta)\hbar q}{\pi m_{||}\rho^2}\sum_{s = \pm}\frac{e^{\pi\nu^{\ell}_{s,2}}}{k^{\ell}_{F,s}}\nonumber\\
& &\times\biggl|B_s^{\ell}e^{i(k^{\ell}_{F,s}\rho-\eta_{\nu^{\ell}_s})}+
C_s^{\ell}e^{-i(k^{\ell}_{F,s}\rho-\eta_{\nu^{\ell}_s})}\biggr|^2\label{jcazimuthal}\\
j_{\varphi}^s(\rho)\!&=&-\frac{2\mu_BE_{\ell}(n-\delta)\hbar}{\pi m_{||}c^2\rho^2}\sum_{s = \pm}\frac{se^{\pi\nu^{\ell}_{s,2}}}{k^{\ell}_{F,s}}\nonumber\\
& &\times\biggl|B_s^{\ell}e^{i(k^{\ell}_{F,s}\rho-\eta_{\nu^{\ell}_s})}+
C_s^{\ell}e^{-i(k^{\ell}_{F,s}\rho-\eta_{\nu^{\ell}_s})}\biggr|^2.\label{jsazimuthal}
\end{eqnarray}
We note that Eq. ({\ref{spincurrent}) implies that the $s=+$ and $s=-$ terms in Eq. (\ref{chargecurrent}) are identical.

When both spin states are  occupied, the wave functions can be normalized to the total particle number  $\overline{n_s}$ for each spin state in the Corbino disk
\begin{eqnarray}
\overline{n^{\ell}}&=&\sum_{s=\pm}\overline{n^{\ell}_s},\label{overlinen}\\
\overline{n^{\ell}_s}
&=&4\frac{e^{\pi\nu^{\ell}_{s,2}}}{k^{\ell}_{F,s}}\int_{\rho_i}^{\rho_o}d\rho\Bigl(|B^{\ell}_s|^2+|C^{\ell}_s|^2\nonumber\\
& &+2{\rm Re}\Bigl[B^{\ell}_sC^{\ell *}_se^{2i(k^{\ell}_{F,s}\rho-\eta_{\nu^{\ell}_s})}\Bigr]\Bigr).\label{overlinens}
\end{eqnarray}

Note that the total charge in the disk is $q\overline{n}$.
 The experimenter must be able to work in both regions of positive and negative $z^{\ell}_{n,s}$.  For $z^{\ell}_{n,s}>0$, $\nu^{\ell}_{s,2}=0$ and $\nu^{\ell}_s=\nu^{\ell}_{s,1}$, which is real.  For $z^{\ell}_{n,s}<0$, $\nu^{\ell}_{s,1}=0$ and $\nu^{\ell}_s=i\nu^{\ell}_{s,2}$, which is imaginary.  The experimenter introduces a potential difference $\Delta v_{\ell}$  in the Corbino disk, as sketched in Figs. 1(b) and 1(c). Then by measuring the voltages from the NM and from the FM electrodes to the electrode on the outer perimeter at $\rho_o$,  separately for $\nu^{\ell}_{s,2}=0$ and for $\nu^{\ell}_{s,2}\ne0$, the experimenter can determine values for $j_{\rho}^c$ between $\rho_i$ and $\rho_{\rm expt}$ and between $\rho_{\rm expt}$ and $\rho_o$, where $\rho_{\rm expt}$ is the radial position(s) of the centrally located NM and FM electrodes.  Since the sheet current is uniform, the overall current density is inversely proportional to $\rho$.  Since there should be no radial spin current, the values obtained from the radial NM and FM electrodes to $\rho_i$ electrode should be the same, as should the values obtained from the radial NM and FM electrodes to $\rho_o$ electrode.  These measurements should provide information of the radial function $D^{\ell}_{s,1}(\rho)$ given by
 \begin{eqnarray}
 D^{\ell}_{s,1}(\rho)&=& \frac{e^{\pi\nu^{\ell}_{s,2}}}{\rho}\biggl(|B^{\ell}_s|^2-|C^{\ell}_s|^2\nonumber\\
& &-\frac{qA_{\ell}}{k^{\ell}_{F,s}}\biggl|B_s^{\ell}e^{i(k^{\ell}_{F,s}\rho-\eta_{\nu^{\ell}_s})}+
C_s^{\ell}e^{-i(k^{\ell}_{F,s}\rho-\eta_{\nu^{\ell}_s})}\biggr|^2\biggr),\nonumber\label{Ds1}\\
& &
 \end{eqnarray}
 which should be the same for $s=+$ and $s=-$.

  Then, by measuring the voltage difference between the two neighboring NM electrodes and between the FM and its neighboring NM electrode, also separately for $\nu^{\ell}_{s,2}=0$ and for $\nu^{\ell}_{s,2}\ne0$ near different $n$ values, the experimenter can infer values for $j^c_{\varphi}$ and $j^s_{\varphi}$, and obtain measurements of the parameters
\begin{eqnarray}
D^{\ell}_{s,2}&=& \frac{(n-\delta) e^{\pi\nu^{\ell}_{s,2}}}{k^{\ell}_{F,s}\rho^2_{\rm expt}}\nonumber\\
& &\times\biggl|B_s^{\ell}e^{i(k^{\ell}_{F,s}\rho_{\rm expt}-\eta_{\nu^{\ell}_s})}+
C_s^{\ell}e^{-i(k^{\ell}_{F,s}\rho_{\rm expt}-\eta_{\nu^{\ell}_s})}\biggr|^2. \label{Ds2}
\end{eqnarray}
The combination of these measured $D^{\ell}_{s,i}$ values would give rise to measured values of $|B^{\ell}_s|$, $|C^{\ell}_s|$, $k^{\ell}_{F,s}$, $\eta_{\nu^{\ell}_s}$, $\nu^{\ell}_{s,2}$, and to the phase difference  $\zeta^{\ell}_s$ given by
\begin{eqnarray}
e^{i\zeta^{\ell}_s}&=&\frac{B^{\ell}_sC^{\ell *}_se^{2i(k^{\ell}_{F,s}\rho_{\rm expt}-\eta_{\nu^{\ell}_s})}}{|B^{\ell}_s||C^{\ell}_s|}
\end{eqnarray}
measured at $\rho_{\rm expt}$.  This second set of experiments at the fixed $\rho_{\rm expt}$ is the most important, and by changing the values of $E_{\ell}$, one can quantify the quantized azimuthal spin current generated by the quantum spin Hall Hamiltonian.

\section*{Acknowledgments}
We acknowledge discussions with Luca Argenti,  Wei Han, Masahiro Ishigami, Abdelkader Kara, Madhab Neupane, and Jingchuan Zhang.
  Work by Q. G. and A. Z. was supported by the National Natural Science Foundation of China through Grant no. 11874083.  A. Z. was also supported by the China Scholarship Council. R. A. K. was partially supported by the U. S. Air Force Office of Scientific Research (AFOSR) LRIR \#18RQCOR100, and the AFRL/SFFP Summer Faculty Fellowship Program provided by AFRL/RQ at WPAFB. Support for T. J. H. and T. J. B. was provided by AFOSR LRIR \#18RQCOR100 and the U. S. Air Force Laboratory - Aerospace System Directorate (AFRL/RQ).\\
 {\bf Author contributions:}  A. Z. and R. A. K. did the theoretical analysis and wrote the paper. T. J. B. designed the experimental setup.  Q. G. and T. J. H. supervised the overall operation, suggested points to demonstrate, and contributed to the final editing of the manuscript.\\

\end{document}